# Room-Temperature Quantum Emission from Boron Nitride Nanotubes

Ke-Sean J.M. Peter[1], Ryan A. Kowalski[1], David A. Curie[2], Dakota Braaten[1], Matthew W. Day[2,3], Denis R. Candido[4,5], Deyu Li[6], Richard F. Haglund[2], Joshua D. Caldwell[1,6,*], Richarda Niemann[6,¥].

1. Interdisciplinary Materials Science Graduate Program, Vanderbilt University, Nashville, Tennessee, 37240
2. Department of Physics and Astronomy, Vanderbilt University, Nashville, Tennessee, 37240
3. Department of Electrical and Computer Engineering, Vanderbilt University, Nashville, Tennessee, 37240
4. Department of Physics and Astronomy, The University of Iowa, Iowa City, Iowa, 52242
5. Applied Mathematical & Computational Sciences, The University of Iowa, Iowa City, Iowa, 52242
6. Department of Mechanical Engineering, Vanderbilt University, Nashville, Tennessee, 37240

* josh.caldwell@vanderbilt.edu

¥ richarda.niemann@vanderbilt.edu

**Abstract**

Quantum information systems such as quantum key distribution and quantum communications require bright, high-purity, stable sources of single photons to act as carriers of quantum information. As such, methods for obtaining single photon emitters (SPEs) in solid-state devices capable of articulating with quantum photonic circuits are a necessary obstacle that must be overcome. Hexagonal boron nitride (hBN) flakes have been extensively studied as a host material for SPEs, though working in two dimensions is accompanied by engineering complexity. However, dimensionality can be reduced through the implementation of boron nitride nanotubes (BNNTs). Here we investigate the potential of BNNTs as room-temperature SPEs using confocal photoluminescence microscopy and Hanbury Brown-Twiss interferometry. Quantum emitters were found in BNNTs spanning the entirety of the visible spectrum, including spectral regions where SPEs within hBN have not been reported. Furthermore, each tube was found to possess multiple SPEs at varying emission wavelengths along the length, exhibiting lifetimes as fast as 0.5 ns, which is about two-fold faster than the fastest hBN SPEs. Additionally, we demonstrate that surface passivation/coating mitigates blinking events, therefore increasing the emitter stability. Thus, BNNTs can function as a single structure supporting a variety of multiwavelength SPE sources, while maintaining a compact radius of ~25 nm, making them ideally suited for deterministic placement and integration within nanophotonic devices such as waveguides and interferometric filters.

## Introduction

Single photon sources are increasingly important building blocks for wide-spread adoption of quantum technologies, such as quantum computing, communication, cryptography, and sensing.[1,2] Individual photons act as carriers of information, with encoding occurring at the quantum level. Existing sources include single atoms,[3] color centers in semiconductors,[4–7] quantum dots,[8] and excitons in transition metal dichalcogenides.[9,10] Individually, these materials host narrow linewidth emitters with each variety encompassing a particular region of the optical spectrum.[10] A single material platform that can support narrowband single photon emitters (SPEs) between the ultraviolet and near-infrared wavelengths offers facile integration into quantum technologies. A suite of defect-based SPEs in hexagonal boron nitride (hBN) have been reported, however, spectral gaps separate the different emitters,[11–14] while there also remains challenges in identifying SPEs with sufficient brightness, short enough lifetimes, and/or active defect spin states. In addition to its 2D sheet structure, boron nitride can take other forms, such as one-dimensional (1D) nanotubes, which also support SPEs that are not as limited in their spectral coverage and provide some increased control for local placement.[15–20]

An ideal emitter for quantum systems should exhibit high spectral and temporal purity, high spectral and intensity stability, and short decay-lifetimes.[21] This combination produces photons that are indistinguishable from each other, an essential component for interference-based tasks such as photonic computing.[22] Moreover, short decay lifetimes and concurrently high brightness enables high bitrates as emission intensity increases.[23] While the emission rate can be artificially enhanced by way of cavity enhancements, strong coupling of an emitter to an on-chip photonic platform requires both spectral and spatial alignment.[24] This has necessitated the development of methodologies that articulate 2D flakes into on-chip photonic systems.[25] To couple light efficiently into a photonic system, the emitter dipole must be precisely aligned with the in-coupling channel in three dimensions.[26] This device manufacturing challenge is compounded by large flakes, such as hBN. Photonic circuit design for on-chip quantum optics is therefore prevented from minimizing undesirable modal dispersion so that the in-coupling of the SPE emission into a corresponding waveguide or on-chip photonic component is maximized irrespective of dipole orientation.[27] This design is further restricted to the photon wavelengths produced by the currently known SPEs, and limited further by those that exhibit sufficient brightness and other properties desired. Moreover, high-performance photonics benefits from the limited spectral dispersion accompanied by narrow linewidths, however, single-wavelength structures are in tension with a desire to spectrally multiplex signals with minimal crosstalk.[28] Finally, although two-dimensional flakes of hBN provide a variety of SPEs with emission wavelengths extending from the deep ultra-violet (UV) to the near-infrared (NIR), there are distinct spectral gaps where emitters have not been identified.[10,29]

Here, we experimentally investigate the optical and quantum properties of boron nitride nanotubes (BNNTs) for use as SPEs in the visible spectral range. As an alternative polymorph of BN, BNNTs present a new one-dimensional paradigm for visible SPE host materials. Confocal photoluminescence (PL) microscopy in conjunction with spectral and temporal analysis techniques were utilized to investigate the photophysical characteristics of photoluminescent sites in a population of BNNTs. We demonstrate that BNNTs host multiple SPEs and provide a rich testbed for quantum optics. In particular, we demonstrate the versatility of BNNTs as spatially localized host materials for high-performance SPEs and improve the spectral and temporal stability. The SPEs in BNNTs span the entirety of the visible spectrum in near continuity with high spectral purity, and sub-nanosecond transition speeds, which represents a substantial improvement over reported lifetimes in hBN SPEs. We use atomic layer deposition to passivate the surface of the BNNTs, mitigating blinking events, increasing emitter stability and enhancing single photon purity.

**Results/Discussion**

Boron nitride can take on a variety of polymorphs, including a cylindrical morphology, aptly called a BNNT (Fig 1a). Multi-layered sheets of van der Waals BN layers are curled inwards to form a continuous loop, while maintaining relative interlayer position (SI 1). This results in a rigid high aspect-ratio structure, within a smaller cross-sectional surface area as compared to the planar phases. High-resolution transmission electron microscopy (TEM) of a BNNT reveals an average interlayer spacing $d \cong 0.33$ nm, which is consistent with that of pristine hBN (Fig 1b).[30] This structure does host a variety of point defect sites visible along the otherwise uniform tube.

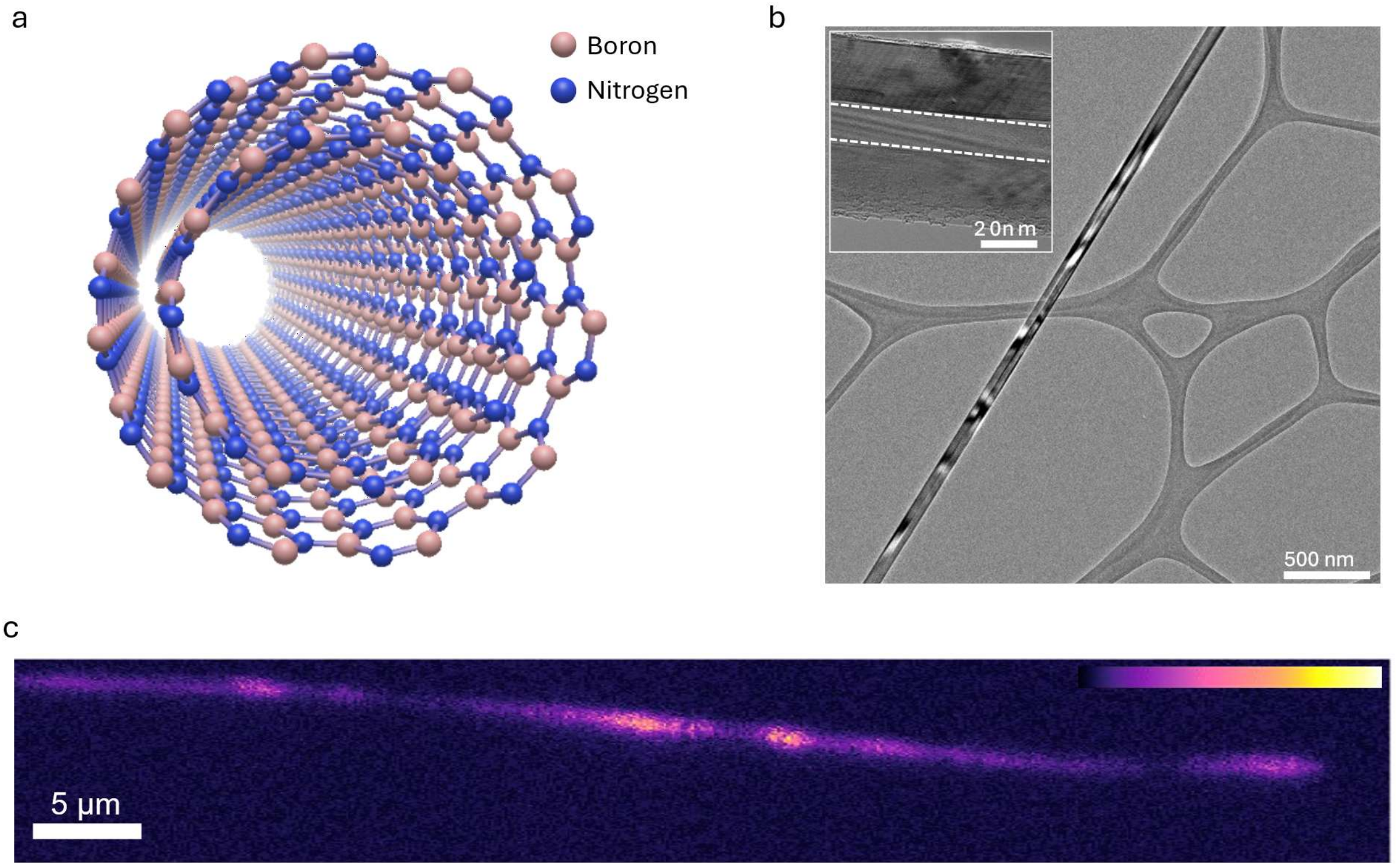

**Figure 1**. Structural analysis of a BNNT. a) Molecular model of bilayer hBN sheets rolling into double-walled nanotube structure. b) Transmission electron micrograph showing a 60 nm diameter multiwalled BNNT. Inset: magnified view to show lattice diffraction. Dashed lines designate inner wall of the nanotube. c) Confocal PL map of the BNNT, showing high emission spots along the axis of the tube under 405 nm continuous wave excitation.

Defects in BNNTs, similar to those in other BN polymorphs, are often photoluminescent at room temperature due to the variety of deep-level transitions available in the wide-bandgap host crystal.[10] Confocal PL microscopy[31] was used to obtain spatial maps of PL intensity of the BNNTs. Numerous defect states contribute to the overall PL of the nanotube, such that most of the tube is visible under 405 nm continuous wave excitation (Fig. 1c). Of these defects, a subset behave as quantum emitters at a variety of spectral wavelengths. Hyperspectral maps of the BNNT extract spatio-spectral information, distinguishing each as its own unique emitter. A broad spectral confocal PL map was collected, showing the total PL intensity coming from the BNNT. While this imaging technique is sufficient for illustrating the relative number of photons being emitted within a region of interest, direct comparisons between emitters ultimately depend on the wavelength-dependent sensitivity of the detector. Spectrally decomposing the hyperspectral map reveals emitters localized at varying sites along the tube, with distinct spatial locations and emission wavelengths. A single BNNT thus possesses several highly localized SPEs (Fig 2a). Spectroscopic analysis of these sites exhibits characteristic features of SPEs spanning the full visible spectrum (Fig. 2b). Emitters with sharp zero-phonon lines (ZPLs) and their corresponding phonon sidebands (PSBs) can be observed across the full visible spectrum (430 – 760 nm, Fig 2b). Some emitters are spatially colocalized and cannot be entirely spectrally or spatially separated, further indicated by second-order autocorrelation ($g^{(2)}$) measurements that are consistent with an antibunching behavior from multiple SPEs (SI 2).

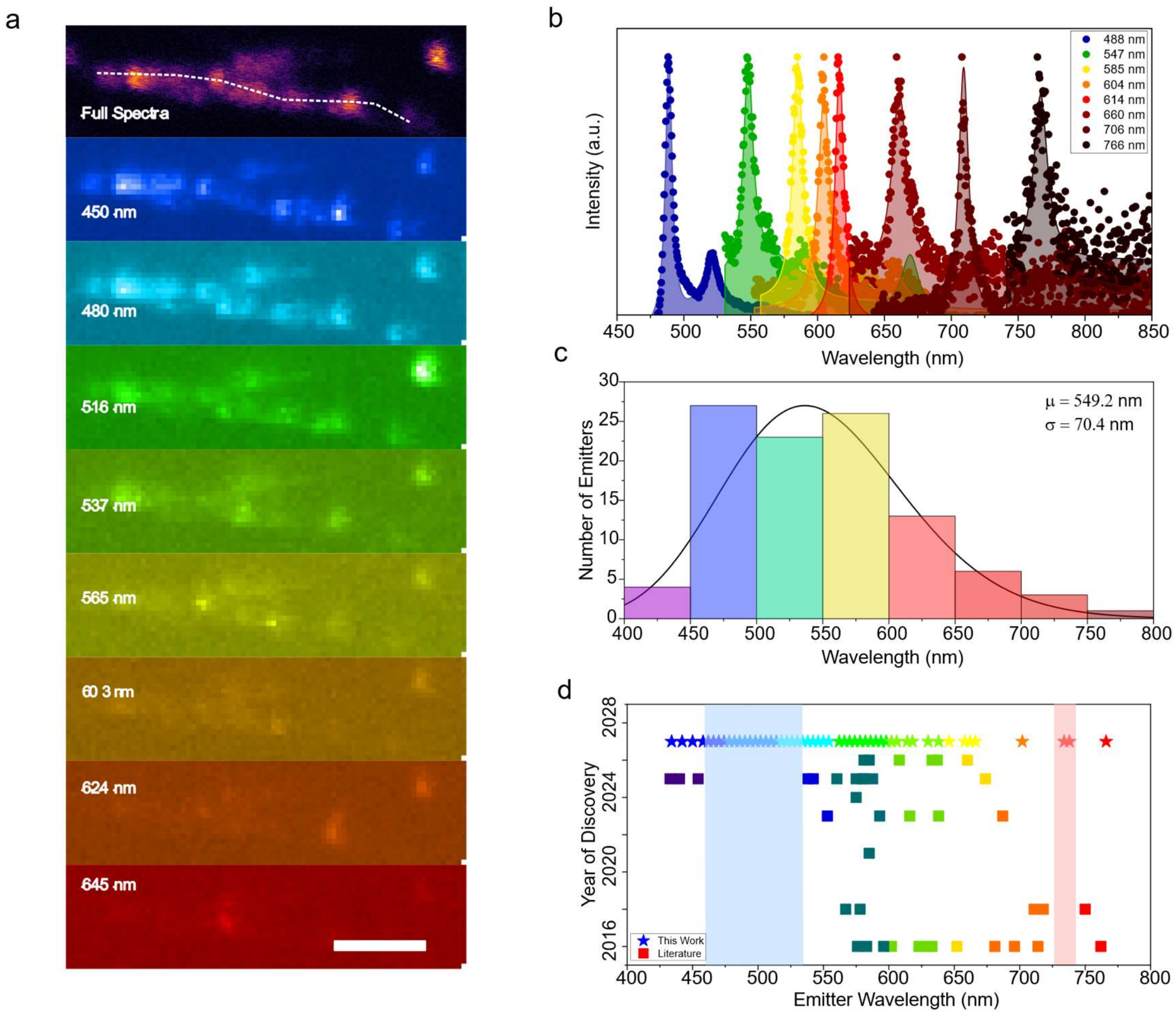

**Figure 2**. Spectral overview of single photon emission in BNNTs over the whole visible spectral range. a) Hyperspectral PL map decomposed to show localization of single photon emitters by wavelength along a BNNT (dashed line). Scalebar 5 μm. b) PL spectra of emitters show characteristic sharp ZPLs and notable PSBs. Symbols show experimental data, shaded regions show cumulative fits. c) Statistical survey of 103 emitters shows larger number of emitters between 450 and 600 nm. d) Chronological discovery of single photon emitters in hBN, as compared to emitters observed in BNNTs.

The frequency with which emitters are observed in a population of BNNTs loosely follows a normal distribution centered at 549 nm (Fig 2c). Approximately 4% of emitters observed exhibited ZPL wavelengths between 400 and 450 nm. Between 450 and 550 nm, ~45% of all emitters in BNNTs are observed, in contrast with the variety of emitters observed in hBN (Fig 2d).[6,14,32–42] This spectral region, highlighted in blue, hosts quantum emitters that are unique to BNNTs. Emitters between 550 and 600 nm, which account for ~25 % of the quantum emitters observed (Fig. 2c), correspond to color centers reported and studied extensively within hBN.[33–35,38–40,42] The frequency that emitters are encountered decreases inversely with wavelength, with an additional 25% of emitters being found between 600 and 800 nm. Nonetheless, BNNTs exhibit full visible spectral coverage of single photon emitters.

We examine one SPE that has a broad ZPL at 706 nm with a FWHM of 10.4 nm, and no notable PSBs (Fig. 3a). The broadening of the ZPL - typically observed in other quantum emitters - is due to thermal contributions to the lattice from the environment at room temperature.[43] The 706 nm emitter is sufficiently spectrally separated from other emission or Raman scattering processes also occurring in the focal spot, such that unperturbed, detailed analysis of its properties is possible.

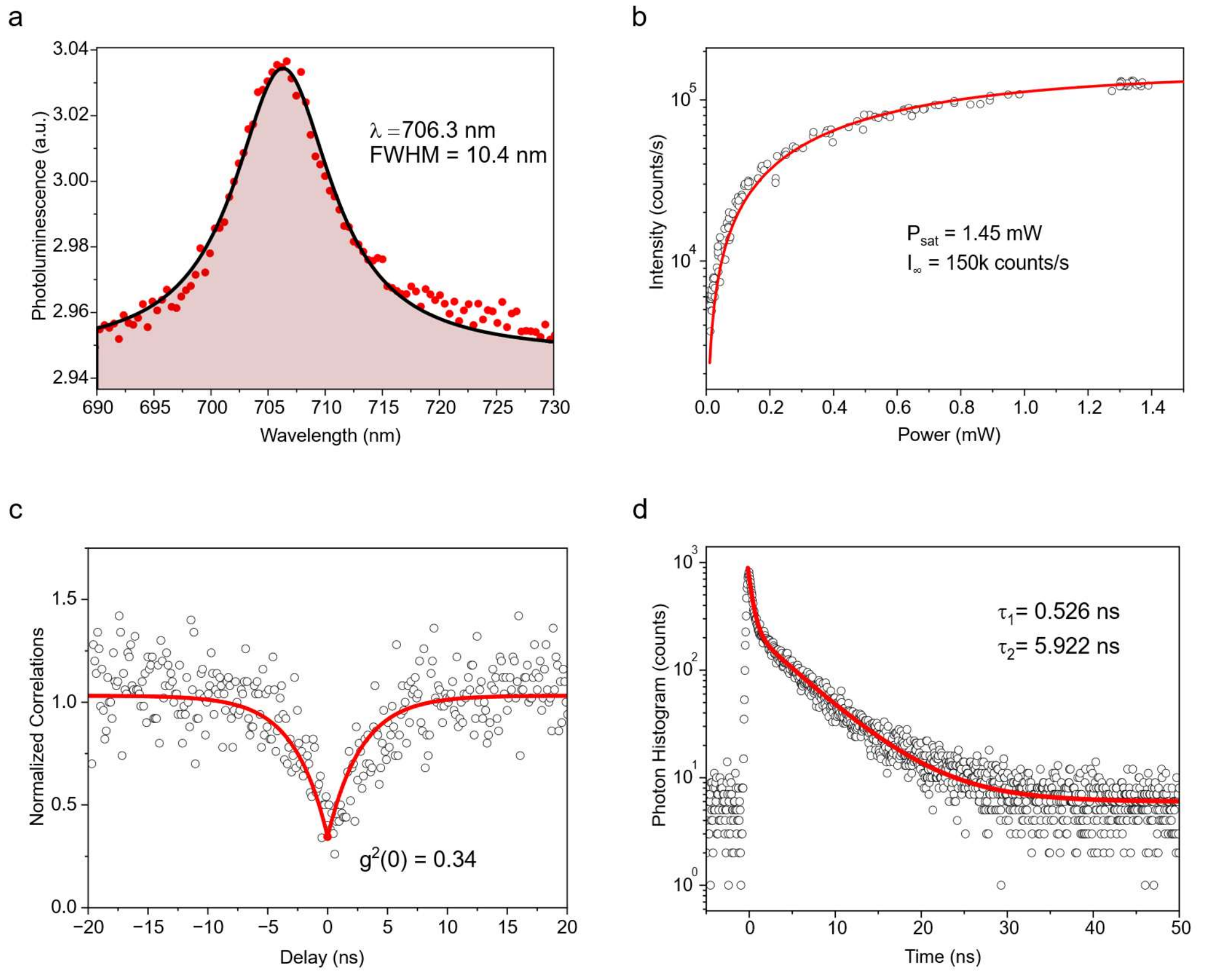

**Figure 3**. Analysis of a spectrally isolated quantum emitter in a BNNT. a) Spectroscopic analysis of emitter shows a ZPL centered at $\lambda$ =706.3 nm, with a linewidth of 10.4 nm. b) PL emission increases with increased excitation power but saturates at 1.45 mW with a maximum emission rate of $\approx 1.5 \times 10^5$ counts/s. c) Second-order correlation for this emitter shows $g^{(2)}(0)$ = 0.34, with a clear antibunching dip being indicative of single photon emission. c) Time-resolved PL lifetime measurement shows two decay lifetimes, a fast decay over $\tau_1 = 0.526$ ns, and a slow decay spanning $\tau_2 = 5.922$ ns.

To probe saturation conditions of the emitter, the excitation power *P* was gradually increased, and PL emission intensity *I* was recorded. The resultant saturation curve (Fig. 3b) was fit using the expression $I = I_{\infty} * (1 - e^{-P_{sat}*P})$ indicating a saturation intensity $I_{\infty}$ of $1.5 \times 10^5$ counts/s at a saturation power $P_{sat}$ of 1.45 mW. A Hanbury Brown-Twiss[44] interferometer was used to perform second-order autocorrelation measurements on the 706 nm emitter below this saturation condition to validate the quantum nature of the emission. Time-correlated photon counts were fit to the equation $g^{(2)}(\tau) = 1 - a \cdot e^{-\frac{|\tau|}{\tau_R}}$, where a is the antibunching amplitude, $\tau_R$ is the decay lifetime for the emitter, and $\tau$ is the time delay.[45] The normalized correlation $g^{(2)}(\tau = 0) = 0.34$, with 100 μW excitation power and no background corrections (Fig. 3c) verifies the quantum nature of the emitter.[44]

The rate at which single photons are generated in a photonic system is related to the emitter lifetime and ultimately defines the bitrate and therefore performance. The decay lifetime of quantum emitters varies depending on the material system, spanning a broad range of timescales, typically ranging from a few picoseconds in quantum dots to several microseconds in silicon color centers.[9,46] To probe the transition lifetime of this emitter, time-resolved PL measurements were performed using a 405 nm pulsed laser with a 10 MHz repetition rate and 100 ps pulse width. A bi-exponential model was used to fit the emitter lifetime decay (Fig. 3d), which typically indicates a multi-level system with both radiative and non-radiative relaxation pathways.[47] The maximum emission rate is inversely proportional to the decay lifetime, resulting in reduced brightness and bitrates for emitters with longer lifetimes.[23] The decay lifetimes in this system were found to be $\tau_1 = 0.53$ ns and $\tau_2 = 5.92$ ns, about half as long as previously reported transition lifetimes $\tau_1, \tau_2$ observed from SPEs in hBN, which typically range from 1 - 10 ns for stable electronic transitions,[6,33,48] resulting in at least an effectively doubled bitrate. The decay lifetimes observed over the population of SPEs in a BNNT range from 0.53 – 8.21 ns (SI 3).

Quantum emitters often suffer from photo-instability, which is observed as a stochastic transition between bright and dark emission states, resulting in fluctuations in intensity.[49] Typically observed in hBN, this phenomenon is attributed to charge traps formed by the introduction of oxygen into the lattice, dangling bonds, or other contamination.[50] Photo-blinking events and the associated spectral instabilities observed in solid state emitters result in the reduction of quantum efficiency,[51] single photon purity,[52] and in turn, reliability for entanglement and other implementations. A known method to reduce photo-blinking events is the isolation of an emitter from the local environment. Passivating the surface mitigates blinking events by saturating dangling bonds and eliminating charge traps near the emitter.[53–55] Here, we employed atomic layer deposition (ALD) to encapsulate the BNNT hosting the 706 nm SPE in 1.5 nm thick film of $Al_2O_3$ (Fig. 4a). As a result, the emission intensity remains stable as a function of time (Fig. 4b), with photo-blinking effects as previously seen in uncoated emitters fully mitigated, with a reduced intensity fluctuation depth.

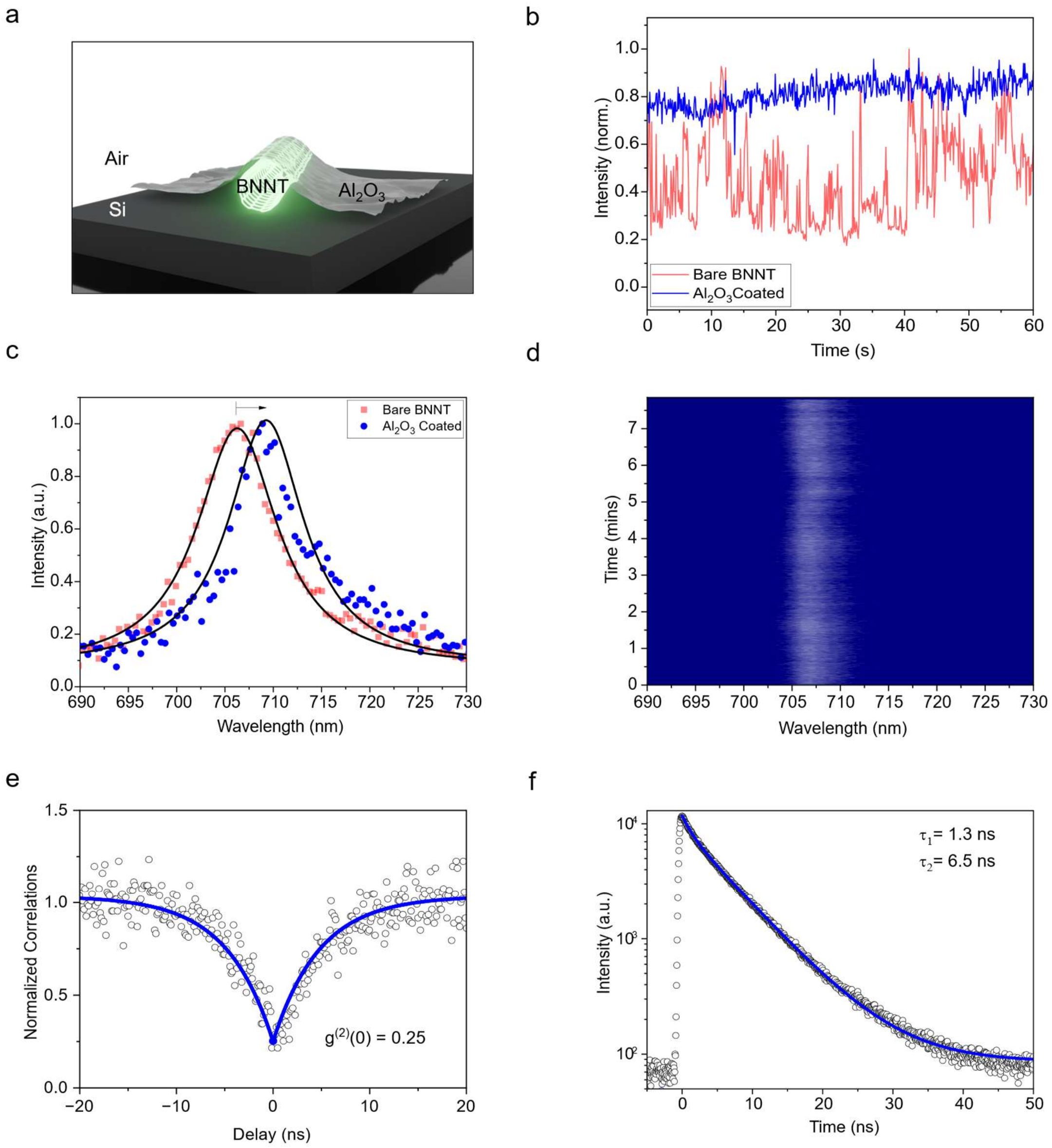

**Figure 4**. Sapphire coating reduces photo-instability of emitters in BNNTs. a) Atomic layer deposition of $Al_2O_3$ conformally coats BNNTs. b) Emission intensity as a function of time shows improved stability (blue), without distinct photo-blinking events as observed in a bare BNNT (red). c) Spectroscopic analysis of emitter shows red-shifting of characteristic ZPL centered from 706 nm (red) to 709 nm (blue) after encapsulation. Solid lines represent Lorentzian fits. d) Temporal evolution shows robust spectral stability centered at emission wavelength. e) Second-order correlation shows improved SPE purity, with $g^{(2)}(0) = 0.25$, confirming single photon emission state. f) Time-resolved PL shows lifetime following a bi-exponential decay, with an increase in decay lifetimes to $\tau_1$=1.3 ns and $\tau_2$= 6.5 ns.

Instabilities associated with photo-blinking are also accompanied by spectral instabilities, and multi-photon emission events. The encapsulated emitter was shown to have undergone a slight red shift in the ZPL from the initial emitting wavelength, from $\lambda_{Bare} = 706$ nm to $\lambda_{ALD} = 709$ nm (Fig 4c). The $Al_2O_3$ coating changes the local electrostatic environment, giving rise to an electric field and hence the observed Stark shift.[55,56] We also find that the spectral stability, i.e., the ZPL variance of the emitter, is improved upon ALD coating and shows excellent consistency over long time periods (Fig. 4d). Under identical excitation conditions, the emitter was selectively probed for single photon emission (Fig. 4e). Autocorrelation measurements ($g^{(2)}(0) = 0.25$) show that the ALD process did not quench or otherwise destroy SPEs housed in the BNNTs, but rather improved single photon purity by suppressing multi-photon emission pathways and chemically isolating the emitter. Local stability induced on the electrostatic environment increases the amount of time the emitter can stay in the excited state.[47] This is experimentally observed as an increase in decay lifetimes of the emitter, to $\tau_1 = 1.3$ ns and $\tau_2 = 6.5$ ns (Fig. 4f). Thus, encapsulation of the BNNTs in $Al_2O_3$ results in spectrally and temporally stable quantum emission. Future work will investigate the role of dielectric coatings on SPE purity and decay lifetimes.

**Conclusion**

Here we demonstrate that BNNTs host SPEs within a confined 1D structure with emission lines observed across the entire visible spectral range. These SPEs exhibit broad ZPLs that can be observed across the entirety of the visible spectrum. This includes regions where SPEs have not been experimentally observed in other BN polymorphs. Transition lifetimes were found to be half that of observed lifetimes in 2D hBN, crossing into the sub-nanosecond regime. One defect that features single photon emission at 706 nm, boasts transition lifetimes that surpass the nano-second limit. The occurrence of SPEs within a BNNT reimagines the dichotomy of either 2D sheets, or 0D quantum dots as the only potential SPE hosts for articulation and integration within photonic systems. A nanotube presents an architecture large enough to be manually manipulated, but with small enough features for precise control of emission ($\pm 1$ tube diameter). Considering that the studied BNNTs are 60 nm in diameter, and double walled nanotubes are only 2 nm in diameter,[57] single axis alignment is effectively all that remains. Furthermore, each tube can host multiple emitters that are independent from each other. This opens the capability for wavelength division multiplexing using a BNNT as a multiwavelength photon source in a single waveguide, drastically increasing bandwidth in quantum communication applications. Furthermore, the quantity of potential emitters is drastically increased, with BNNTs hosting emitters between 450 nm and 550 nm previously unseen in 2D BN systems. Current state-of-the-art SPEs promise to build the next generation of photonic quantum computing, sensors, and cryptographies, but their necessary operational conditions limit the progression of the field.[9,58] SPEs in BN do not require cryogenic conditions, or ultrahigh vacuum, but scale commercial fabrication has not yet been realized. BNNTs can circumvent the fabrication issue probabilistically. SPEs in BNNTs are more than capable of filling a niche in probability-based quantum systems as on-chip embeddable sources of single photons enhanced by conformal dielectric coatings.

## Methods

*Sample Preparation*

Boron nitride nanotubes (BNNTs) were purchased from NanoIntegris, then dispersed in ethanol by placing a small quantity of the as-received material into a vial and sonicating for 60 seconds. The resulting suspension was then drop-cast onto a patterned silicon substrate. After the ethanol evaporated, the deposited BNNTs were suitable for locating individual nanotubes and performing subsequent measurements.

*Measuring and analysing electron micrographs*

High-resolution TEM were collected from BNNTs on carbon coated mesh copper grids. The FEI Tecnai G2 Osiris S/TEM was used under an accelerating voltage of 200 kV, with the primary image collection facilitated by a Gatan CCD and FluScreen. Image analysis was performed in the commercial image processing software ImageJ.

*Optical Characterization*

Emitters were surveyed using a modified WITEC Alpha SNOM confocal Raman microscope. A variable CW/pulsed 405 nm laser (Thorlabs GSL41A) was used for excitation and spatial mapping of emitters. Plasma lines were filtered using a bandpass filter (405nm MaxLine® laser clean-up filter, Semrock). A linear polarizer and half-wave plate (Thorlabs) were used to set the incident polarization state to maximize emission. Excitation and emission were focused and collected using a high NA 100X objective lens (Zeiss, NA = 0.9). The emitted light was filtered using a 415 nm dichroic mirror, a variable shortpass filter (Thorlabs FESHV1) and a variable longpass filter (Thorlabs FELHV1) before being coupled into a spectrometer (Acton SpectraPro 2300i, Princeton Instruments) or to two photon counters (Excelitas SPCM) in a Hanbury Brown-Twiss configuration. The photon counts at both detectors are recorded and correlated using an autocorrelator (Time Tagger Ultra, Swabian Instruments). These correlated counts are used for validating single photon emission or determining emitter lifetimes. In measuring time-resolved PL, the excitation source was configured to 100 ps pulse width, at a 10MHz repetition rate to determine PL lifetimes.